\documentclass[twocolumn,preprintnumbers,amsmath,amssymb,prl]{revtex4-1}
\usepackage{graphicx}
\usepackage{dcolumn}
\usepackage{bm}
\usepackage{color}

\begin{document}
\title{Crystallization Dynamics on Curved Surfaces}

\author{Nicol\'as A. Garc\'ia$^{1}$}
\author{Richard A. Register$^{2}$}
\author{Daniel A. Vega$^{1}$}
\email{lgomez@uns.edu.ar, dvega@uns.edu.ar}
\author{Leopoldo R. G\'omez$^{1}$}
\email{lgomez@uns.edu.ar, dvega@uns.edu.ar}

\affiliation{$^1$ Department of Physics, Universidad Nacional del
Sur - IFISUR
- CONICET, 8000 Bah\'ia Blanca, Argentina\\
$^2$ Department of Chemical and Biological Engineering, Princeton
University, Princeton, New Jersey 08544, USA}

\date{\today}

\begin{abstract}
{We study the evolution from a liquid to a crystal phase in
two-dimensional curved space.  At early times, while crystal seeds
grow preferentially in regions of low curvature, the lattice
frustration produced in regions with high curvature is rapidly
relaxed through isolated defects. Further relaxation involves a
mechanism of crystal growth and defect annihilation where regions
with high curvature act as sinks for the diffusion of domain
walls. The pinning of grain boundaries at regions of low curvature
leads to the formation of a metastable structure of defects,
characterized by asymptotically slow dynamics of ordering and
activation energies dictated by the largest curvatures of the
system. These glassy-like ordering dynamics may completely inhibit
the appearance of the ground state structures. }
\end{abstract}

\maketitle

\section{Introduction}

The evolution from a liquid to a crystal structure is one of the
oldest problems in condensed matter, with wide interest in both
basic science and technology \cite{Chaikin}. In three-dimensional
space the mechanism of crystallization is still a matter of
ongoing research. Since in general the seeds for nucleation do not
have the same symmetry as the equilibrium structure, the early
process of crystallization is a competition among different
intermediate metastable states \cite{Chaikin}-\cite{GomezVega}. On
the other hand, in 2D flat space the physics behind
crystallization is much simpler because there is no frustration to
nucleating the equilibrium structure, such that there is no need
for intermediate precursors \cite{VegaGomez}. However,
two-dimensional (2D) crystals deposited on curved surfaces come
back to complexity. Here the underlying curvature locally
frustrates the formation of the crystal lattice, modifying both
equilibrium structures \cite{Harris1970}-\cite{IrvineVitelli} and
their dynamical properties \cite{BauschNatMat},
\cite{IrvineNatMat}.

Curved crystalline structures are ubiquitous in nature. For
example, they can be found in viral capsids, insect eyes, pollen
grains, and radiolaria. During the last decade these crystals have
attracted the interest of different communities because of the
richness associated with the coupling between geometry, structure,
and functionality. Recently, curved crystals have been obtained in
a controlled fashion by the use of colloidal matter
\cite{BauschScience}, \cite{Irvine}, \cite{BauschNatMat},
\cite{IrvineNatMat}. Other self-assembled systems with great
potential to develop such structures are block copolymers and
liquid crystals \cite{HexemerThesis}-\cite{Shi}. In curved
crystals, defects can be a feature of the fundamental
(equilibrium) state. Depending on the substrate's topology and
curvature, defects can be required to reduce lattice distortions
and to satisfy topological constraints. Thus, from a condensed
matter perspective, the presence of curvature in ordered phases
appears as an opportunity for accurate control of the density and
location of topological defects \cite{Nelson}, \cite{Stellacci}.

Although theoretical and experimental work has led to a
substantial advance in the knowledge of equilibrium structures and
features, the out-of-equilibrium dynamics leading to the formation
of curved crystals, highly relevant for technological applications
like defect functionalization engineering or soft lithography,
remain almost unexplored.

In this work we use a free energy functional that includes
competing interactions to describe the large-scale dissipative
dynamics of crystallization in systems residing on curved
backgrounds. This simple model captures the essential features of
crystallization over diffusive time scales and provides a clear
picture about the complex coupling between geometry and
crystallinity. We focus on the long time dynamics of the system
which is governed by the formation, interaction and annihilation
of topological defects.

\section{Model and Analysis}

The process of crystallization in curved space can be described by
the following free energy functional:
\begin{equation}
\begin{split}
F&=\int  d^2r \, \sqrt{g} \, [W(\psi)+\frac{D}{2} \, g^{\alpha
\beta} \, \partial_\alpha \psi(\textbf{r}) \, \partial_\beta \,
\psi(\textbf{r})]
\\- &\frac{b}{2}\int\int d^2r \, d^2r' \, \sqrt{g} \,
\sqrt{g'}
\,G(\textbf{r}-\textbf{r}')\,\psi(\textbf{r})\,\psi(\textbf{r}')
\end{split}
\end{equation}
Here $\psi$ is an order parameter related to fluctuations of the
density from the average, $W(\psi)=-\gamma \, \psi^2+v \,
\psi^3+u\, \psi^4$ is a double-well potential that below the
critical temperature presents  two minima, $D$ is a penalty to
form interfaces, $\sqrt{g}$ is the determinant of the metric
$g_{\alpha \beta}$ of the curved substrate, and
$G(\textbf{r}-\textbf{r}')$ is a Green's function which takes into
account long-range interactions leading to the formation of a
local hexagonal structure. Equation 1 is the covariant
generalization of models widely used to study properties of
systems with competing interactions with direct applications to a
large collection of condensed systems: block copolymers, crystals,
magnetic multilayer compounds, Rayleigh-Benard convection
patterns, and doped Mott insulators \cite{OhtaKawasaki},
\cite{SeulAndelman}, among others. Close to the critical line, Eq.
1 reduces to the Landau-Brazovsky free energy functional which has
been employed for years in the understanding of the liquid to
crystal transition in Euclidean space.

The dynamics leading to crystallization can be analyzed by
considering the temporal evolution of a liquid phase quenched
below the critical temperature, through the evolution equation:
\begin{equation}
\frac{\partial \psi}{\partial t}=M\nabla_{LB}^{2}\{\frac{\delta
F}{\delta \psi}\}
\end{equation}

Here $M$ is a phenomenological mobility coefficient, and
$\nabla^2_{LB}$ is the Laplace-Beltrami operator which reduces to
the classical Laplacian in flat geometries (see the appendix for
details on the evolution equation and numerical methods).

In this work we focus on the crystallization of hexagonal systems
residing on sinusoidal substrates of different curvatures (Figure
1a), characterized by an amplitude $A$ and a wavelength $L$ larger
than the lattice constant in the crystal.  In general, for every
regular point $P$ on the surface there are two tangent circles
with maximal and minimal radii of curvature $R_1$ and $R_2$,
respectively \cite{Oneill}. The Gaussian curvature $K$ at $P$ is
then defined as $K=\kappa_1 \kappa_2$, where $\kappa_i=1/R_{i,
i=1,2}$ are the principal curvatures. The Gaussian curvature
represents the intrinsic curvature of the surface. In the geometry
studied here it takes on negative (positive) values in the
neighborhood of the saddles (crests or valleys). Regions of zero
Gaussian curvature correspond to points resembling the flat
Euclidean plane. Note that for these surfaces the curvature is
symmetrically distributed (see Figure 1b), such that the
integrated curvature is zero. Thus, this system is topologically
equivalent to the plane, such that no defects are geometrically
required to form the crystal lattice; defect structures, if any,
must arise only from energetic considerations.

\begin{center}
\begin{figure}[t]
\includegraphics[width=7.5 cm]{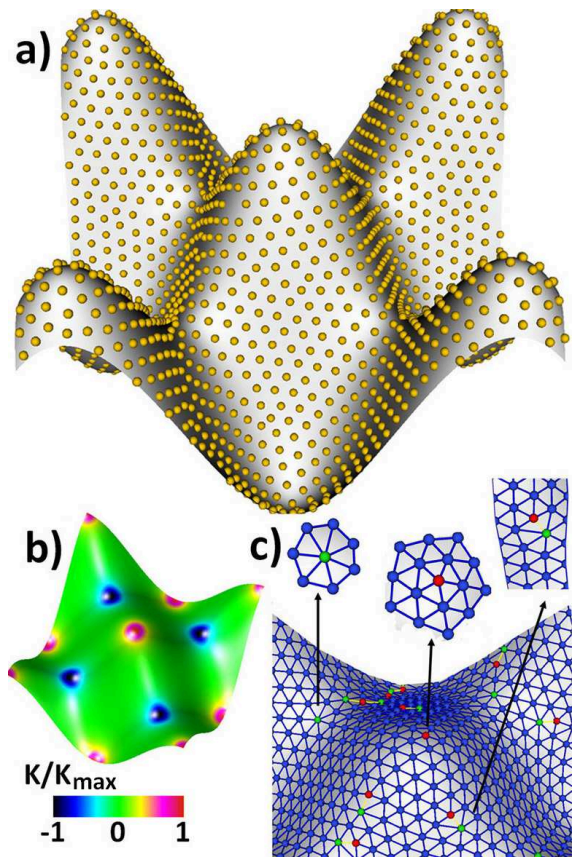}\caption{a) Relaxed configuration of a hexagonal crystal lying on a sinusoidal substrate.
b) Distribution of Gaussian curvature. Here $K_{max}$ is the
maximum curvature. c) Delaunay triangulations are used to identify
the topological defects like negative (left) and positive (center)
disclinations, and dislocations (right).}
\end{figure}
\end{center}

Upon crystallization, the positions of the particles can be
determined through  the local maxima in the order parameter
function $\psi (\textbf{r})$. Once the particles' positions are
known, the temporal evolution of the coordination number,
particles' first neighbors, degree of crystallinity, pair
correlation function, and topological defects can be analyzed
through Delaunay triangulations (see the appendix for
supplementary information about the method used to obtain Delaunay
triangulations on curved surfaces).

The most common defects of a hexagonal flat crystal are given by
particles whose coordination number differs from six (Fig. 1c)
\cite{Chaikin}. Five- and seven-coordinated particles, named
positive and negative disclinations, respectively, are deeply
involved in the high temperature behavior of 2D crystals and also
unavoidably created during a symmetry-breaking liquid-to-crystal
transition. In flat systems these topological defects are highly
energetic because they produce large distortions in the crystal
lattice. Consequently, in flat crystals disclinations are not
found isolated but are arranged in dipoles, known as dislocations
(Fig. 1c). On the contrary, on curved substrates the disclinations
can help to screen out the geometric frustration induced by the
substrate's geometry, reducing the elastic distortions generated
by the curvature \cite{Harris1970}-\cite{IrvineVitelli}. Thus, in
curved crystals the disclinations are \emph{not} necessarily
defects as they can belong to the equilibrium state of the
crystal.

\begin{figure*}[t]
\begin{center}
\includegraphics[width=17 cm]{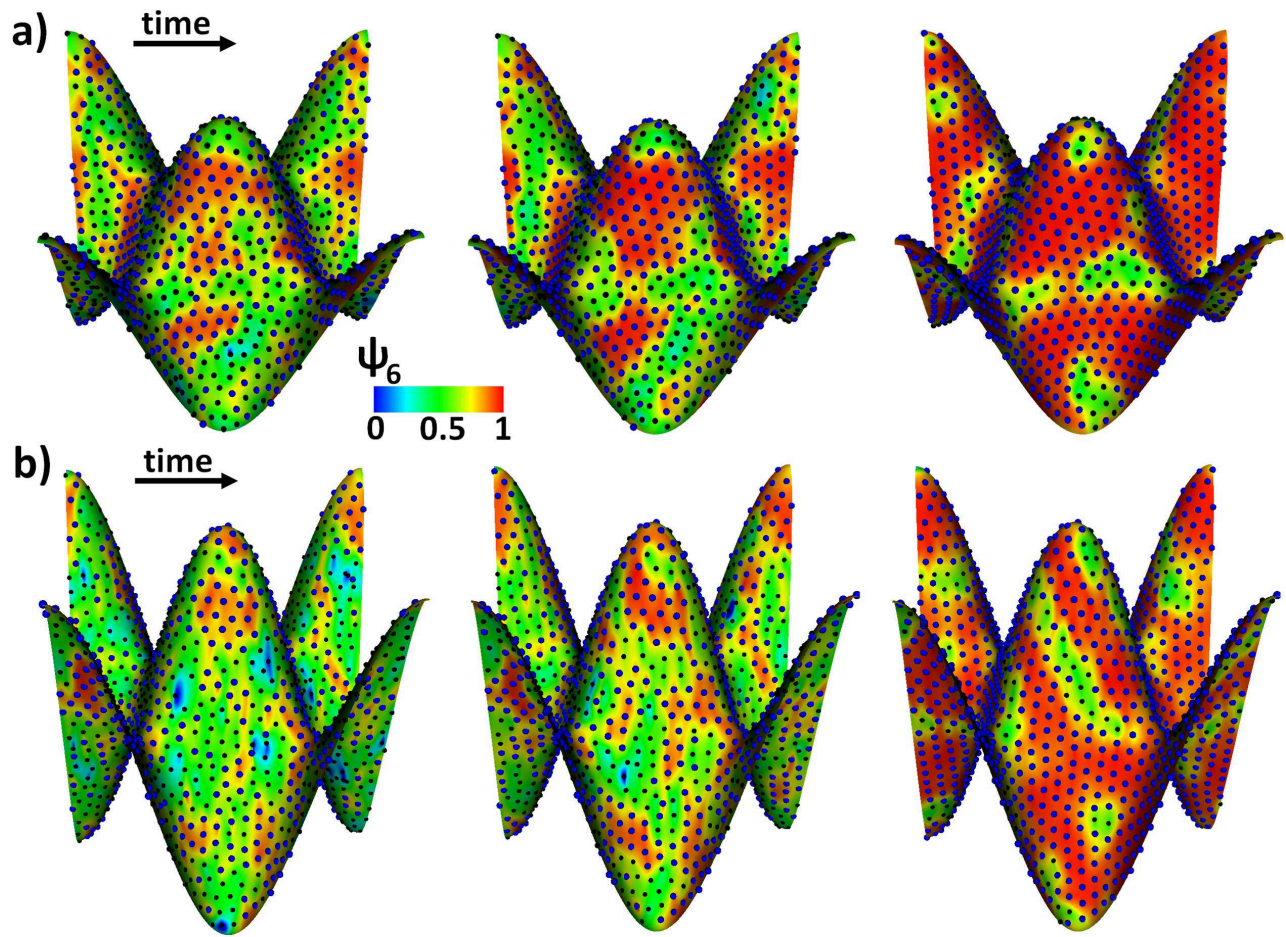} \caption{Formation of crystal domains from an initial liquid phase at
curvatures $K_{max}\,a^2=0.64$ (a), and $K_{max}\,a^2=1.18$ (b).
Horizontal panels corresponds to same evolution times ($t=1500$
top, $t=3000$ middle, and $t=15000$ bottom). While crystal
particles are drawn as big blue spheres, liquid-like particles are
drawn as smaller black spheres. Note the formation of the early
hexagonal nuclei in the flattest regions of the substrate (left
and middle panels). For large curvatures the hexagonal domains
start to develop rather elongated shapes (compare the red regions
in the right panels of a and b).}
\end{center}
\end{figure*}

\section{Results}

\textbf{Crystal patch formation:} The formation of the crystal
order can be analyzed through the bond-orientational order
parameter $\psi_6^j=\frac{1}{Z_j} \sum_k exp(6i\theta_{jk})$ for
each particle $j$, where the sum runs over the $Z_j$ nearest
neighbors of particle $j$, and $\theta_{jk}$ is the angle between
the $j-k$ bond and a fixed direction. This order parameter takes a
value $\psi_6^j=0$ for a liquid particle, and $\psi_6^j=1$ for a
crystal particle.

Sequences 2a and 2b show the early crystallization process from an
initial liquid, on two differently curved substrates (in Fig. 2a
$K_{max}\,a^2=0.64$, and in Fig 2b $K_{max}\,a^2=1.18$, with $a$
being the lattice constant). To better visualize the crystal
regions, here we have superimposed the bond order parameter maps
$\psi_6$ with the particles.  Note that in both cases the
formation of hexagonal order (red regions) starts in the flatter
regions of the substrate. In flat systems the initial crystal
seeds of a liquid-solid transition are randomly distributed
throughout the system \cite{Chaikin}, \cite{VegaGomez}. On the
contrary, here the curvature breaks the isotropy of space, and the
formation of hexagonal patches is favored on regions where the
Gaussian curvature is small. This is a consequence of the
geometrical frustration created by the curvature, where the
geometry produces distortions that increase the strain energy in
the lattice, inhibiting the formation of perfectly hexagonal
patches in regions of high curvature.

This phenomenon can be clearly observed by studying the
distribution of the crystal order parameter $\psi_6$ as a function
of the local value of the Gaussian curvature $K$. Figures 3a and
3b show $\psi_6$ for early and long times during the
crystallization on substrates of increasing curvature. Since the
geometric frustration is reduced by disclinations that locally
reduce crystallinity, by increasing the maximum curvature
$K_{max}$ the crystallinity becomes sharply peaked around the
flatter regions ($K\sim 0$) at early and long times. Note that as
the curvature of the system is increased, the confinement of
crystal seeds to regions of low curvature is stronger.

The global process of crystallization can be tracked through the
fraction of particles involved in  crystals, $f_C=\sum_j \psi_6^j/
\sum_j$. Figure 3c shows the temporal evolution of $f_C$. This
plot shows that the dynamics become slower and the fraction of
crystallized particles decreases by increasing the curvature.
Thus, the curvature not only frustrates the formation of hexagonal
domains in regions of high curvature, but also affects the
dynamics towards the equilibrium state.

\begin{figure}[t]
\begin{center}
\includegraphics[width=8.5 cm]{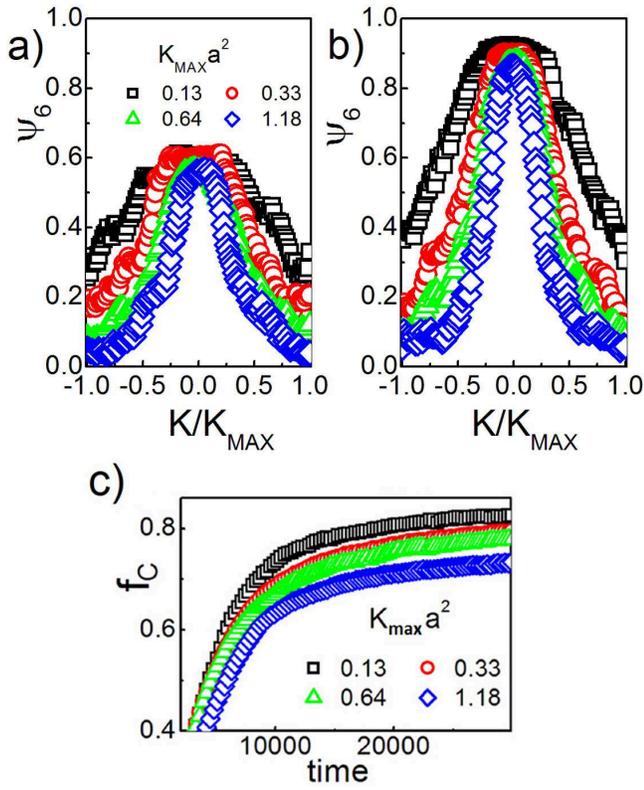} \caption{Distribuion of the order parameter $\psi_6$ as a function
of the Gaussian curvature  $K$ at times $t=3000$ (a) and $t=20000$
(b). The curves corresponds to different substrates (curvatures
indicated in the figure). c) Temporal evolution of the crystal
fraction $f_C$. Here $a$ is the lattice constant.}
\end{center}
\end{figure}

\begin{center}
\begin{figure}[b]
\includegraphics[width=8.5 cm]{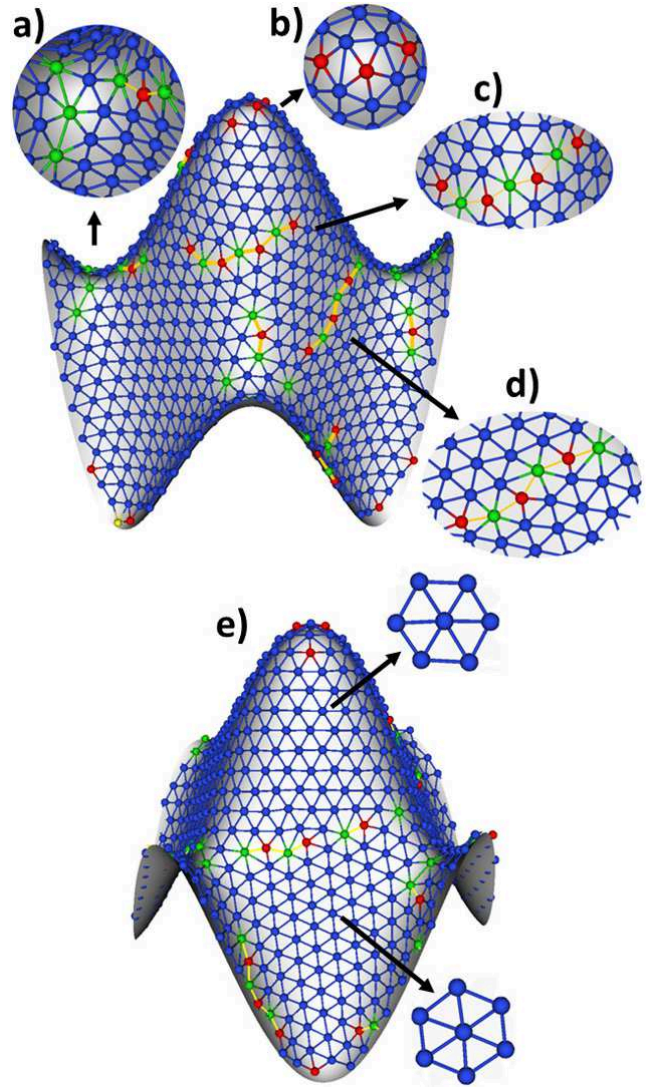} \caption{Negative (panel a) and positive
(panel b) disclinations located at saddles and bumps,
respectively, reduce the geometric frustration produced in regions
with high curvature. Scars (c) and pleats (d) delimit different
curved hexagonal crystals. e) Domain structure observed at long
times. Here grain boundaries become pinned to regions of low
curvature, slowing the kinetics of ordering. }
\end{figure}
\end{center}

\textbf{Defect structures:} We observed that the early crystalline
seeds grow through the addition of new hexagonally packed
particles, invading the rest of the substrate. However, the growth
is not homogeneous and the crystals grow preferentially faster
along regions of low curvature. At intermediate time scales, where
the system remains far from equilibrium, we observed defect
configurations similar to those observed in curved colloidal
crystals (Fig. 4). Here the different structures of defects that
participate in the dynamics of crystallization are dislocations,
pleats (linear arrays of dislocations with variable distance
between them \cite{Irvine}), free disclinations, and scars (linear
arrays of dislocations and disclinations with a net topological
charge) \cite{BauschScience}. During crystallization in flat
space, the domain structure is dominated by triple points (regions
where three grains meet) \cite{GomezVegaprl}, \cite{Pezzutti};
isolated disclinations are absent \cite{VegaPRE},
\cite{Harrison2}. However, due to curvature, here the triple
points become unstable and the mechanism of domain growth produces
scars and pleats (see Fig. 4) decorating the interface between
grains having a large orientational mismatch, while free
disclinations are rapidly stabilized in the highly curved regions
in order to screen the geometric potential \cite{Nelson}.

\textbf{Pair correlations:} The relaxation towards the ordered
crystal requires the annihilation of the excess defects, while
keeping the total topological charge equal to zero. A standard
tool to study the ordering process is the pair correlation
function $g(r)$, giving the probability of finding a particle at
distance $r$ away from a reference particle \cite{Rubinstein} (see
Appendix C). For a perfectly ordered crystal, $g(r)$ is a periodic
array of $\delta$ functions whose positions and heights are
determined by the lattice order. Since disorder strongly affects
the peak structure, it has been previously found that $g(r)$ can
be employed to study the degree of ordering. Figure 5 shows $g(r)$
for a highly curved substrate ($K_{max}a^2\sim 1$) at early and
long times. The process of ordering is evidenced by the sharpening
of higher-order peaks in $g(r)$, which correspond to the peak
structure found for a hexagonal lattice (peaks located at
$r/a=1,\sqrt{3},2,\sqrt{7},3,\ldots$). When comparing $g(r)$ at a
fixed time, we observe that the amplitude of the first peak in
$g(r)$ systematically decreases upon increasing the maximum
curvature $K_{max}$. However, unlike other systems where the
degree of order affects the distribution of peak heights, here we
observe that the ratio between peak heights is relatively
insensitive to the curvature (inset of Fig. 5). Thus, although at
high curvatures the system contains a number of isolated
disclinations and other defect structures, their highly energetic
elastic distortions are deeply relaxed by the geometry.

\begin{center}
\begin{figure}[t]
\includegraphics[width=8.5 cm]{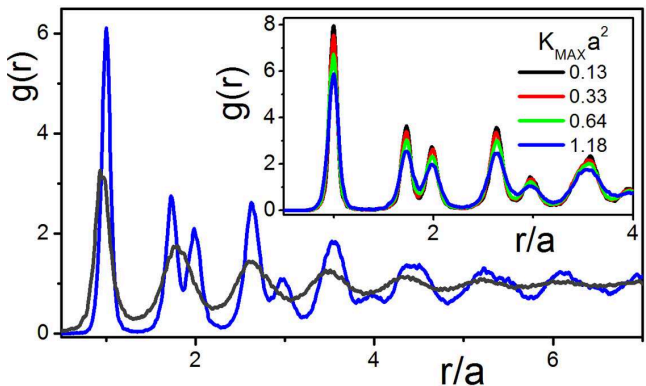} \caption{Pair correlations
$g(r)$ at short (gray) and long (blue) times for a highly curved
substrate ($K_{max}a^2=1.18$). Inset: $g(r)$ for systems of
increasing curvature ($t=3\times10^5$).}
\end{figure}
\end{center}

\textbf{Defect evolution and correlation length:} The coarsening
dynamics can also be elucidated by tracking the motion and
distribution of topological defects. At early times, the disorder
in the lattice screens out the local curvature, producing a
roughly random distribution of defects throughout the system
\cite{Tarjus}.  At this early stage the dynamics are relatively
unaffected by the curvature and thus, the process of defect
annihilation can be expected to follow  ``line-tension-driven"
Allen-Cahn dynamics \cite{GomezVegaprl}, \cite{AllenCahn}.  As
time proceeds, most defects condense along domain walls in pleats
and scars, while their dynamics of diffusion and annihilation are
deeply influenced by both isolated disclinations \cite{deWit},
\cite{Harris1971} and curvature. The sequence of Fig. 6a shows the
mechanism of ordering most often observed in regions of high
curvature. Here the partially screened strain field of the
disclinations located at the bumps acts as a sink for linear
arrays of defects, that diffuse and annihilate at these highly
curved regions.

The analysis of the defect structures and the annihilation
mechanism reveals that in addition to the local traps produced by
regions of positive (negative) curvature for the motion of
positive (negative) disclinations, a slow \emph{glassy-like}
ordering results from the pinning of linear arrays of defects to
regions of low curvature.  Figure 4e shows a typical pinning of a
domain wall connecting negative disclinations, located at saddle
points. We found that the deep traps produced by the regions of
low curvature generate very stable domain structures that
dramatically slow down dynamics by freezing the ordering kinetics
\cite{Berthier}-\cite{Safran}. Note however, that this slowing
down is not produced by the formation of a glassy phase, but
rather related to an arrested state, which is unable to reach
equilibrium. Although in Euclidean space the slow dynamics of
systems with competing interactions has for years been related to
the formation of glasses, recent results indicate that the
dynamics in the two cases would be intrinsically different
\cite{Berthier}, \cite{Reichman}.

This slowing down in the dynamics can be tracked through the time
evolution of the areal density of dislocations $\rho_{ds}$ (Figure
6b). It can be observed that in agreement with the results shown
in Fig. 3c, at short times the dynamics are relatively insensitive
to the curvature while a clear coupling with the geometry appears
at intermediate and long times. In this regime, upon increasing
the maximum curvature, the rate of defect annihilation decreases
and larger contents of defects are involved in the relaxation
process.

\begin{center}
\begin{figure*}[t]
\includegraphics[width=12 cm]{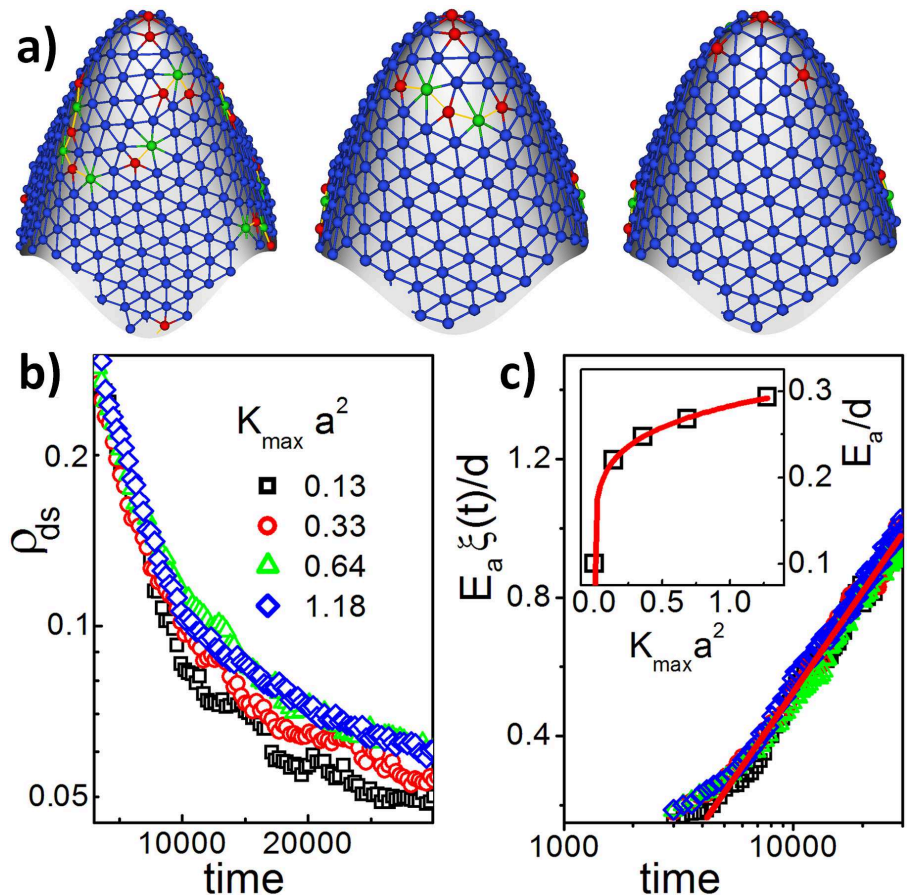}
\caption{a) Linear arrays of defects decorating domain walls
diffuse to regions of high curvature where they are absorbed by
isolated disclinations. Panels b) and c) show $\rho_{ds}$ and
$\xi$ as functions of time, for systems with different underlying
curvatures. Note that upon normalization by $E_a$,  $\xi$
asymptotically follows a similar slow logarithmic evolution,
irrespective of $K_{max}$. The inset of panel c) shows $E_a$ as a
function of $K_{max}$. The value of $E_a$ found in flat systems
($K_{max}=0$) has also been included. The line represents a
power-law fit, $E_a \sim K_{max}^{0.23}$.}
\end{figure*}
\end{center}

Since defect structures mainly decorate domain walls, a
characteristic length scale $\xi$, defined in terms of $\rho_{ds}$
as $\xi \sim d/\rho_{ds}$, provides a measure of the average
domain size \cite{GomezVegaprl}-\cite{Harrison2}. Here $d \sim 2a$
is the average distance between dislocations along a domain wall.
In coarsening systems where a free energy barrier $U_a$ is
involved in the growth of the domains, the rate of change of the
correlation length takes the form $\frac{d\xi}{dt}=
exp{(-\frac{U_a}{kT})}$ \cite{GomezVegaprl}. Since here the energy
excess is produced mainly by the domain walls, we have $U_a/kT =
E_a\,\xi/d$, where $E_a$ is a free energy parameter characterizing
the strength of the barrier.  Solving for $\xi$,  the asymptotic
behavior of $\xi$ becomes $\xi(t)\sim\frac{d}{E_a} \ln
\frac{E_a}{d}t$. Figure 6c shows the temporal evolution of $E_a
\xi(t)/d$ for substrates with different curvature $K_{max}$. In
agreement with an activated mechanism, here we find that in the
long-time regime, $\xi$ is consistent with a logarithmic
dependence on time. We also found that the mean-square
displacement of the particles is consistent with a logarithmic
behavior. The inset of Fig. 6c also shows the dependence of
$E_{a}$ as a function of the maximum curvature of the substrate
$K_{max}$. The activation energy for a flat system has also been
included for comparison. In agreement with the qualitative
observations, the activation energy grows continuously with the
maximum curvature, indicating that the pinning of defects becomes
stronger for larger curvatures. However, the dependence on
curvature is relatively weak, with $E_{a}$ following a power law
$E_{a}\sim K_{max}^{\eta}$ with a relatively small exponent ($\eta
=  0.23$).

\section{Conclusions}

In this work we have studied how the mechanisms of formation of a
two-dimensional crystal phase are modified by the presence of
curvature.  We observed that order is triggered earlier on the
flattest regions of the substrate, where the frustration in the
lattice due to the curvature is smaller. The frustration to form
the crystal  induced by the geometry of the substrate  can be
partially reduced by different topological defect structures.

The ordering mechanism of Curved polycrystals show ordering
mechanisms somewhat similar to those of their flat counterparts,
modified by curvature. As the coupling between varying curvature
and crystal order induces the pinning of grain boundaries,
glassy-like dynamics arise which may completely inhibit the
appearance of the ground state structure.  These dynamic effects
should be taken into account not only at the onset of a phase
transition during crystallization or melting, but also in the
design of applications that require well-ordered self-assembled
structures, like surfaces functionalized with ordered arrays of
topological defects.

\section{Acknowledgements}

This work was supported by the National Research Council of
Argentina, CONICET, ANPCyT, Universidad Nacional del Sur and the
National Science Foundation MRSEC Program through the Princeton
Center for Complex Materials (DMR-0819860).

\appendix

\section{Appendix A: Evolution Equation}

By combining the free energy functional (Eq. 1) with the
relaxational equation (Eq. 2), we obtain a partial differential
equation describing the evolution of the order parameter during
crystallization:
\begin{equation}
\frac{\partial \psi}{\partial t}=M\nabla_{LB}^{2}
[f(\psi)-D\nabla_{LB}^{2}\psi]-Mb \psi
\end{equation}
where $f(\psi)=dW(\psi)/d\psi=-\gamma \psi +v \psi^2+u \psi^3$,
and $\nabla_{LB}^{2}$ is the Laplace-Beltrami operator which in
its general form can be written as \cite{Struik}:
\begin{equation}
\nabla_{LB}^{2}\equiv \frac{1}{\sqrt{g}}\frac{\partial}{\partial
x^i}(g^{ij}\,\sqrt{g} \, \frac{\partial}{\partial x^j})
\end{equation}
Here $g$ is the determinant of the metric of the substrate
$g_{ij}=\frac{\partial \textbf{R}}{\partial x^i}\cdot
\frac{\partial \textbf{R}}{\partial x^j}$, and
$g_{ik}g^{kj}=\delta_i^j$.

In the present work the evolution equation for the order parameter
is numerically solved on the sinusoidal geometry:
\begin{equation}
\textbf{R}(x^1,x^2)=x^1 \, \textbf{i}+x^2 \, \textbf{j}+A
\,cos(2\pi x^1/L) \, cos(2\pi x^2/L) \,\textbf{k}\nonumber
\end{equation}
We have implemented a finite difference algorithm, centered in
space and forward in time, on a 1024x1024 square grid with
periodic boundary conditions. In order to avoid any numerical
coupling with the underlying lattice, the parameters of the free
energy and the number of lattice points of the numerical grid are
chosen such that the particle diameter is represented by at least
$10$ lattice points. The initial liquid phase is modelled by
random fluctuations in the order parameter $\psi$. The evolution
of particles during crystallization is followed by the
identification and tracking of the maxima of the order parameter
\cite{Crocker}.

\begin{center}
\begin{figure}[t]
\includegraphics[width=8 cm]{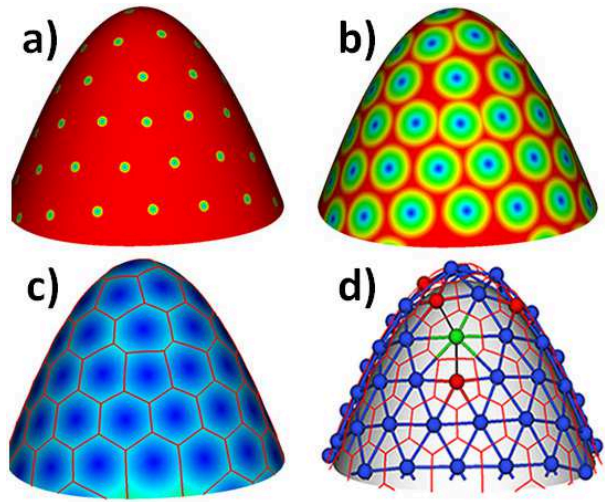} \caption{Delaunay triangulation construction on curved surfaces
by the use of a fast front marching technique. The propagation of
fronts from each particle (a and b) allows the determination of
the first neighbors, Voronoi diagram (c), and Delaunay
triangulation and topological defects (d).}
\end{figure}
\end{center}

\section{Appendix B: Delaunay Triangulation on arbitrarily curved surfaces}

As pointed out above, the dominant features of the system  can be
analyzed by the use of Delaunay triangulation \cite{Harrison2}.
Through the triangulation, it is possible to obtain structural
parameters such as the particle's first neighbors, degree of
crystallinity, or topological defects
\cite{GomezVegaprl}-\cite{Harrison2}. Once the positions of the
particles are determined, the analysis starts by calculating each
particle's first neighbors. To do that, we implemented the Fast
Marching algorithm \cite{Sethian}. Here a propagating front
starting from each particle is obtained by solving the evolution
equation (Figs. 7a and 7b).
\begin{equation}
\frac{\partial \Lambda}{\partial t}+\vec{u} \cdot \vec{\nabla}
\Lambda = 0
\end{equation}
where $\vec{u}$ is the propagation velocity ($|\vec{u}|=1$). The
resulting function $\Lambda$ gives the distance to the initial
point, such that the geodesic distance between two arbitrary
points on the surface is easily determined.

With the geodesic distances between particles we construct the
Voronoi diagrams and Delaunay triangulations and also locate the
particles' first neighbors and determine their coordination
numbers $Z$ (Figs. 7c and 7d). This allows the identification of
topological defects in the structure.

\begin{center}
\begin{figure}[t]
\includegraphics[width=8.5 cm]{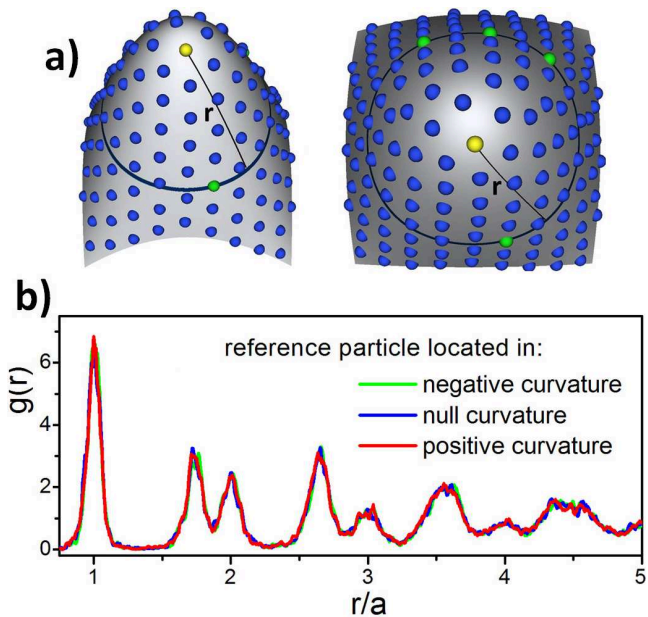} \caption{a) Scheme showing the calculation of pair correlations in curved space.
Here the reference particle is in yellow, and the particles at a
distance $r$ are in green. b) Pair correlations of a crystal-like
structure calculated using particles located in regions of
negative (green), null (blue), and positive (red) curvatures.}
\end{figure}
\end{center}

\section{Appendix C: Pair correlations}

In curved space, the pair correlation function $g(r)$ can be
generalized as the average number of particles with geodesic
distances between $r$ and $r+dr$ \cite{Rubinstein}. Figure 8a
shows a scheme of the calculation of $g(r)$ through the use of
geodesic circles. Here the spatial correlations between the
particle's locations produce peaks in $g(r)$ at characteristic
distances $r^*$ related to the lattice structure (Fig. 8b).

Contrary to $3D$ or flat $2D$ systems, or to homogeneously curved
systems (spheres or pseudospheres), on sinusoidal substrates one
should question whether it is proper to simply average $g(r)$
calculated for different particles, because the varying curvature
breaks the homogeneity of space. However, as shown in Fig. 8b, we
have only found small differences in $g(r)$ when calculated with
reference particles located in curved or flat regions. Thus, the
average of $g(r)$ for the different particles represents a good
measure of the state of order of the system. The regularity in the
lattice structure, producing similar pair correlations independent
of the reference particle's location, comes from the ability to
pack topological defects in the most curved regions of the
substrate.

\end{document}